# Adaptive DDoS attack detection method based on multiple-kernel learning

Jieren Cheng [1, 2 ,4,]   Chen Zhang[1, *],   Xiangyan Tang[1],   Victor S. Sheng[3], Zhe Dong[1], Junqi Li[1], Jing Chen[1]

[1]College of Information Science & Technology, Hainan University, Haikou 570228, China.

[2]State Key Laboratory of Marine Resource Utilization in South China Sea, Haikou 570228, China.

[3]Department of Computer Science, University of Central Arkansas, Conway, AR 72035, USA.

[4]Key Laboratory of Inernet Information Retrieval of Hainan PROVince, Hainan University, Haikou 570228, China;

Correspondence should be addressed to Chen Zhang; 314848554@qq.com.

Distributed denial of service (DDoS) attacks have caused huge economic losses to society. They have become one of the main threats to Internet security. Most of the current detection methods based on a single feature and fixed model parameters cannot effectively detect early DDoS attacks in cloud and big data environment. In this paper, an adaptive DDoS attack detection method (ADADM) based on multiple kernel learning (MKL) is proposed. Based on the burstiness of DDoS attack flow, the distribution of addresses and the interactivity of communication, we define five features to describe the network flow characteristic. Based on the ensemble learning framework, the weight of each dimension is adaptively adjusted by increasing the inter-class mean with a gradient ascent and reducing the intra-class variance with a gradient descent, and the classifier is established to identify an early DDoS attack by training simple multiple kernel learning (SMKL) models with two characteristics including inter-class mean squared difference growth (M-SMKL) and intra-class variance descent (S-SMKL). The sliding window mechanism is used to coordinate the S-SMKL and M-SMKL to detect the early DDoS attack. The experimental results indicate that this method can detect DDoS attacks early and accurately.

## 1. Introduction

In recent years, the security of computer networks, chips, virtual networks and mobile devices has been widely concerned [1-3]. As an important platform for information exchange, computer network security has attracted much attention. In the security of computer network, Distributed denial of service (DDoS) attack is yet to be settled in a long time. DDoS is a traditional network attack method. It controls a large number of zombie machines sending a large number of invalid network request packets to a target host. It consumes and meaninglessly occupies the resources of the server, causing normal users to be unable to use the normal services provided by the target host [4]. Although the DDoS attack mode is simpler, its destruction power to the network is far more than other network attacks.

Moreover, this traditional attack method in recent years can still cause great damage to the Internet, and the frequency of launch, loss caused, complexity of DDoS, diversity of DDoS and difficulty of defense have increased more than before [5]. In June 2016, an ordinary U.S. jewelry online sales website was flooded with 35,000 HTTP requests (spam requests) per second, making the site unable to provide normal services. In October, DynDNS, which provides dynamic DNS services in the United States, was subject to large-scale DDoS attacks, resulting in access problems for multiple websites using DynDNS services, including GitHub, Twitter, Airbnb, Reddit, Freshbooks, Heroku, SoundCloud, Spotify, and Shopify. Twitter has even appeared in nearly 24 hours with a zero-visit situation. The reason why DDoS attacks have such a great destructive power is that DDoS uses a large number of zombie machines to launch attacks on a certain target. Each zombie machine has powerful computing capability. Through the massive distributed processing capabilities of zombie machines, it is easy for a server to no longer have the ability to provide services to normal users [6]. On the other hand, DDoS attacks are easy to implement. Unlike other network attacks, DDoS attacks require only a large number of zombie machines and a small amount of network security knowledge to launch an effective attack. This easy-to-grasp network attack method makes the DDoS attack more powerful.

At present, under the traditional network environment, methods for defense against DDoS attacks mainly include attack detection and attack response [7]. DDoS Attack detection is based on attack signatures, congestion patterns, protocols, and source addresses as an important basis for detecting attacks, thereby establishing an effective detection mechanism. The detection model can be roughly divided into two categories: misuse-based detection and anomaly-based detection. Misuse-based detection is a technique based on feature-matching algorithms. It matches the collected and extracted user behavior features with the known feature database of DDoS attacks to identify whether an attack has occurred. Anomaly-based detection is adopted by monitoring systems. By establishing the target system and the user's normal behavior model, the monitoring systems can determine whether the states of the system and the user's activities deviate from the normal profile and can judge whether there is an attack. The attack response is to properly filter or limit the network traffic after the DDoS attack is initiated. The attack traffic to the attack target host is reduced as much as possible to mitigate the influence of the denial of a service attack.

With the rise of cloud computing technologies and software-defined networking (SDN) concepts, DDoS attack detection based on cloud computing environments and software-defined networks has received widespread attention [8, 9]. As a new computing model, cloud computing has powerful distributed computing capabilities, massive storage capabilities, and diverse service capabilities [10, 11]. It has become an important means of solving big data problems [12]. Therefore, establishing a cloud platform system is a necessary measure to effectively ensure cloud computing's reliability, stability and security [13-15].

In recent years, machine learning has been applied to the field of security [16]. The method of constructing an attack detection model using machine learning has been widely used [17, 18]. The machine learning method plays an important role in the traditional network environment, the cloud environment and software-defined network architecture. The reason is that the machine learning method can deeply mine the important information hidden behind the data and combine prior knowledge to discriminate and predict new data [19]. Therefore, compared with traditional detection methods, machine learning methods can exhibit better detection accuracy [20-24]. In the above analysis of defense measures, it can be known that the traditional network environment, cloud environment and software-defined network architecture all involve attack detection for the defense mechanism of DDoS. Therefore, studying the use of machine learning methods to identify DDoS attacks is of great significance. However, the data generated by the DDoS attack is often burst and diverse, and the background traffic size also has a greater impact on the detection model, thereby reducing the model's detection accuracy.

To solve the above problems, we propose a multiple-kernel learning DDoS attack detection method. The method uses the algorithm to extract five features and combines two multiple-kernel learning models with the adaptive feature weights to recognize attack flows and normal flows. For further improving the accuracy of DDoS attack detection, a sliding window mechanism is employed to coordinate two multiple-kernel learning models treating the detection results. Experiments show that our method can better distinguish DDoS attack flow from normal flow and can detect DDoS attacks earlier.

## 2. Related work

DDoS attacks can cause tremendous damage to a network and often subject the attacked party to great economic losses. This is one of the main ways that hackers initiate cyberattacks.

To reduce the damage of DDoS attacks, researchers have proposed a large number of attack detection methods in recent years. According to the application scenario, these methods can be divided into three categories: the detection method in the conventional network environment, the detection method in the cloud environment, and the detection method in the software-defined network (SDN) environment.

(1) The conventional network environment refers to the Internet environment generally established on the Internet based on an open system interconnect reference model (OSI). In this regard, Saied et al. proposed a method for detecting known and unknown DDoS attacks using artificial neural networks [25]. Bhuyan et al. proposed an empirical evaluation method for the measurement of low-rate and high-rate DDoS attack detection information [26]. Tan et al. proposed a DDoS attack detection method based on multivariate correlation analysis [27]. Yu et al. proposed a DDoS attack detection method based on the traffic correlation coefficient [28]. Wang et al. conducted an in-depth analysis of the characteristics of DDoS botnets [29]. Kumar and others used the Jpcap API to monitor and analyze DDoS attacks [30]. Khundrakpam et al. proposed an application-layer DDoS attack detection method combining entropy and an artificial neural network [31].

(2) The cloud environment refers to the network service platform with cloud computing as the core technology. In this regard, Karnwal et al. proposed a defense method for XML DDoS and HTTP DDoS attacks under cloud computing platforms [34]; Sahi et al. proposed the check and defense method for TCP-flood DDoS attacks in the cloud environment [35]. Rukavitsyn et al. proposed a self-learning DDoS attack detection method in the cloud environment [36].

(3) Software-defined network refers to a new network architecture that adopts OpenFlow as the communication protocol and specifies the router as well as switch data exchange rules through the controller [37]. In this regard, Ashraf use machine-learning detection software to define DDoS attacks under the network [38]. Mihai–Gabriel proposed an intelligent elastic risk assessment method based on the neural network and risk theory in the SDN environment [39]. Yan et al. proposed an effective controller scheduling method to reduce DDoS attacks in software-defined networks [40]. Chin et al. proposed a DDoS flood attack method for selective detection of packets under SDN [41]. Dayal et al analyzed the behavioral characteristics of DDoS attacks under SDN [42]. Ye et al proposes a method of using SVM to detect DDoS attacks under the SDN environment [43]. Except the above detection methods used to ensure the security of the system, some efficient cryptography techniques can be applied to achieve privacy of the system [44-47].

In summary, the core issue of DDoS attack detection research is the construction of feature extraction and classification models. The attack detection methods in the above three environments can effectively detect DDoS attacks corresponding to the environment. However, in the detection of early DDoS attack, these defense methods do not have a good detection effect. In addition, most of these methods use a single feature and do not consider the impact of multidimensional features on the classifier. Therefore, an adaptive DDoS attack detection method is proposed in this paper. Firstly, we design the algorithms to extract five features. Secondly, through an ensemble learning framework, the five features are used to train two multi-kernel learning models and obtain the adaptive feature weights with gradient method. Finally, the sliding window mechanism is used to coordinate the two models to improve the detection accuracy.

## 3. DDoS attack feature extraction

### 3.1 Analysis of DDoS attack behavior

In the cloud environment, the botnets of DDoS attacks have distributed characteristics. Each zombie machine has the ability to independently calculate, send and process data packets, and the source IP address of the packets can also be forged. The advantage of these DDoS attacks make defense more difficult. However, under the background of time series, the characteristics of data packets generated by DDoS attacks are still quite different from those of normal users. The difference is reflected in the following three aspects:

(1) Asymmetry

DDoS attack is often caused by multiple zombie hosts sending a large number of packets to a host without the host's response. These useless packets quickly consumes the host's service resources so that the host can no longer provide services to other users. With this feature, the DDoS attack behavior is such that there are a large amount of packets sent to the host form the zombie hosts, and there are no or a small amount of packets sent to the zombie hosts form the host. The IP data packet often presents a situation in which multiple-source IP addresses point to the same or several destination IP addresses, which is expressed as the asymmetry of the source IP as well as the destination IP in sending and receiving.

(2) Interactivity

Assuming that there are A (zombie host) and B (attacked host). When an attack occurs, there are two main communication ways as follows: (1) A sends packets to B (denoted as A→B); (2) A and B send packets to each other (denoted as A⇄B). And the packet amount sent with the way (A→B) is much more than those sent with the way (A⇄B). Therefore, the interactivity of DDoS attack flow has different states in communication direction and amount compared with normal flow.

(3) Distribution

According to the characteristics of DDoS attack, when an attack occurs, the number of the hosts that launch the attack is much larger than that of the attacked hosts. And the number of the source IP address is much larger than that of the destination IP address, so that the source address and the destination address have different distribution characteristics. In addition, because DDoS attacks generate useless requests, so compared to normal flows, the host ports of accessed by the attack requests are more dispersed. Therefore, the distribution of the ports is different in normal flows and attack flows.

Due to the limited ability of a single feature to express data, it cannot fully reflect the characteristics of the DDoS attack. Therefore, to effectively express the characteristics of the DDoS attack, this paper selects five feature extraction methods based on the above characteristics. That is, the address correlation degree (ACD) combines the traffic burstiness, flow asymmetry, and source IP address distribution of DDoS attack; the IP flow features value (FFV) exploits the asymmetry of attack flows and the distribution of source IP addresses; the IP flow's interaction behavior feature (IBF) uses the different interactivity between normal flows and attack flows on the network; the IP flow multi-feature fusion (MFF) exploits the different behavioral characteristics of normal flows as well as DDoS attack flows and integrates the multiple characteristics of DDoS attack flows; the IP flow address half interaction anomaly degree (HIAD) focuses on the characteristics of the aggregated attack flows that are mixture of a large number of normal background flows. In order to make the feature richer in representation, we refer to several articles and combine the five feature extraction algorithms, besides removing the less impactful parameters to form a multidimensional feature for DDoS attack detection. [46–52].

**3.2 DDoS attack feature extraction**

In the cloud environment, assume that network flow $F$ is as follows: $<(t_1,s_1,d_1,p_1),(t_2,s_2,d_2,p_2),......,(t_n,s_n,d_n,p_n)>$ in a certain unit of time, where $t_i, s_i, d_i, p_i$ denotes the time, source IP address, destination IP address and the port of the $i(i=1,2,.....,n)$-th data packet, respectively. All data packets which contain source IP address $A_i$ and destination IP address $A_j$ are denoted as class $SD(A_i, A_j)$. All data packets with source IP address $A_i$ are denoted as class $IPS(A_i)$. All data packets with destination IP address $A_j$ are denoted as class $IPD(A_j)$. The packets with source IP address $A_i$ which exist in the class $IPS(A_i)$ and class $IPD(A_i)$ are denoted as $IF(A_i)$. The packets with source IP address $A_i$ which exist in class $IPS(A_i)$ and do not exist class $IPD(A_i)$ are denoted as $SH(A_i)$. The number of the different ports in $SH(A_i)$ is denoted as $Port(SH(A_i))$. The packets with the destination IP address $A_i$ which do not exist in class $IPS(A_i)$ and exist in class $IPD(A_i)$ are denoted as $DH(A_i)$. The number of the different ports in $DH(A_i)$ is denoted as $Port(DH(A_i))$.

**Definition 1:** If there are different destination IP addresses $A_j$ and $A_k$, making classes $SD(A_i, A_j)$ and $SD(A_i, A_k)$ both non-null, then delete the class where all source IP address $A_i$ packets reside.

Assume that the last remaining classes are denoted as $ACS_1, ACS_2, ....., ACS_m$, and are statistically calculated to gain the ACD. The detailed formulation is as follows:

$$ACD_F = \sum_{i=1}^{m} W(ACS_i) \quad (1)$$

In this part $W(ACS_i) = \theta_1 Port(ACS_i) + (1-\theta_1)Packet(ACS_i)(0<\theta_1<1)$ where $Port(ACS_i)$ is the number of different ports in class $ACS_i$, $Packet(ACS_i)$ is the number of data packets in class $ACS_i$, and $\theta_1$ is the weighted value.

**Definition 2:** If all the packets whose destination IP address is $A_j$ form the unique class $SD(A_i, A_j)$, delete the class where the packet with the destination IP address is $A_j$.

Assume that the last remaining classes are denoted as $SDS_1, SDS_2, ....., SDS_l$, and all packets in these remaining classes with the destination IP address $A_j$ are denoted as $SDD(A_j)$, and all the classes are denoted as $SDD_1, SDD_2, ....., SDD_m$. The FFV is defined as follow:

$$FFV_F = (\sum_{i=1}^{m} CIP(SDD_i) - m) \quad (2)$$

$CIP(SDD_i)$ in formula (2) is presented as follows:

$$CIP(SDD_i) = Num(SDD_i) + \theta_2 \sum_{j=1}^{Num(SDD_i)} OA(Pack(A_j)) + (1-\theta_2)(OB(Port(SDD_i)) - 1) \quad (3).$$

In this equation, $0 \leq \theta_2 \leq 1$, $Num(SDD_i)$ is the number of different source IP address in $SDD_i$;

$$OA(Pack(A_j)) = \begin{cases} Pack(A_j) & Pack(A_j)/\Delta t > \theta_3 \\ 0 & Pack(A_j)/\Delta t \leq \theta_3 \end{cases}$$

, $Pack(A_j)$ is the number of source IP addresses $A_j$ in $SDD_i$, and $\theta_3$ is the threshold of the number of packets:

$$OB(Port(SDD_i)) = \begin{cases} Port(SDD_i) & Port(SDD_i)/\Delta t > \theta_4 \\ 0 & Port(SDD_i)/\Delta t \leq \theta_4 \end{cases}$$

, $Port(SDD_i)$ is the number of different destination ports in $SDD_i$, $\theta_4$ is the threshold of the number of ports, and $\Delta t$ is the sampling time.

**Definition 3:** Assume that the IF flow is $IF_1, IF_2, ......, IF_M$, the SH class is denoted as $SH_1, SH_2, .....SH_S$, and the DH class is denoted as $DH_1, DH_2, ......, DH_M$; then, define IBF as follows:

$$IBF = \frac{1}{M+1}(|S-D| + \sum_{i=1}^{S} over(Port(SH_i)) + \sum_{i=1}^{D} over(Port(DH_i))) \quad (4)$$

$$over(x) = \begin{cases} x & x/\Delta t > \theta_5 \\ 0 & x/\Delta t \leq \theta_5 \end{cases}$$, where $\theta_5$ is the threshold of the amount of port. $M$ in formula (4) is the number of IF flows within $\Delta t$, $|S-D|$ is the absolute value of the difference value between the number of source IP addresses and the number of destination IP addresses for all SH and DH flows in $\Delta t$.

**Definition 4:** Assume that the resulting SD classes are $SD_1, SD_2, \cdots SD_l$ and IF classes are $IF_1, IF_2, \cdots IF_M$. The number of packets of source IP address $A_i$ in class $IF_i$ is denoted as $Sn_i$, where $i = 1, 2, ..., M$; the number of packets of all interworking flow classes is denoted as SN; and the source semi-interactive flow class is denoted as $SH_1, SH_2, \cdots SH_S$. The number of different port in class $SH_i$ is denoted as $Port(SH_i)$, where $i = 1, 2, ..., S$; the destination semi-interactive class is denoted as $DH_1, DH_2, \cdots DH_D$; and the number of different port in class $DH_i$ is denoted as

$Port(DH_i)$, where $i$ = 1, 2, ..., $D$.

The weighted value of all packets in SH class is defined as follows:

$$Weight_{SH} = \sum_{i=1}^{s} oversh(Packet(SH_i)) \quad (5)$$

The weighted value of all packets in SD classes is defined as follows:

$$Weight_{SD} = \sum_{i=1}^{L} oversd(Packet(SD_i)) \quad (6)$$

The weighted value of the number of packets of network flow F in unit time T is as follows:

$$Weight_{packet} = flag(Weight_{SD})Weight_{SD} + Weight_{SD} \quad (7)$$

In these equations,

$$oversh(x) = \begin{cases} x, x/\Delta t > \theta_6 \\ 0, x/\Delta t \leq \theta_6 \end{cases}$$

$$oversd(x) = \begin{cases} x, x/\Delta t > \theta_7 \\ 0, x/\Delta t \leq \theta_7 \end{cases}$$

$$flag(x) = \begin{cases} 0, x > 0 \\ 1, x = 0 \end{cases}, \Delta t \text{ is sampling time,}$$

$\theta_6$ and $\theta_7$ are SH-type packet number abnormality thresholds; $Packet(SD_i)$ is the number of packets in $SD_i$, $I$ = 1, 2, ..., $n$. The weighted value of the number of different ports in the SH and DH classes is as follows:

$$Weight_{port} = \sum_{i=1}^{S} overp(Port(SH_i)) + \sum_{j=1}^{D} overp(Port(DH_j)) \quad (8)$$

where, $overp(x) = \begin{cases} x, x/\Delta t > \theta_8 \\ 0, x/\Delta t \leq \theta_8 \end{cases}$, $\Delta t$ is sampling time, $\theta_8$ is the SH-type port number abnormality threshold.

In this part we define the MFF is as follows:

$$MFF_F = \frac{S + Weight_{port} + Weight_{packet}}{M+1} \quad (9)$$

where $f(x) = \begin{cases} x, x \geq 1 \\ 1, x \leq 1 \end{cases}$.

**Definition 5:** The number of SH flows with different source IP addresses and the same destination IP address $A_i$ is denoted as $hn_i$. The SH class with the same destination IP address $A_i$ flow is denoted as $HSD(hn_i, A_i)$, where $i$ = 1, 2 ,..., $n$.

Assume that all HSD classes are $HSD_1, HSD_2, \cdots HSD_k$, and the number of different destination port in the class $HSD_i$ is expressed as $Port(HSD_i)$, where $i$ = 1, 2 ,..., $k$.

The HIAD is defined as follows:

$$HIAD_F = \left( \sum_{i=1}^{k} (hn_i + weight(Port(HSD_I))) \right) \quad (10)$$

In eq. (10), $weight(x) = \begin{cases} x, x/\Delta t > \theta_9 \\ 0, x/\Delta t \leq \theta_9 \end{cases}$,

$\Delta t$ is sampling time, and $\theta_9$ is the threshold for different destination port.

## 4. The DDoS attack detection model

The establishment of an attack detection model is an important part of the whole detection process. Based on the behavior of DDoS attack, we extract *ACD*, *IBF*, *MFF*, *HIAD* and *FFV* features to express the inherent rules of attack flows. The disadvantages of the current DDoS attack detection models are summarized as follows: (1) some models highly depend on the selection of kernel function; (2) some models require data with highly stable value; (3) some models can only fit linear rules, but DDoS attack can generate linearly inseparable data due to abrupt, unstable and stochastic characteristics.

Considering that the multiple-kernel learning model has a low requirement for data stability and can be used for nonlinear fitting, furthermore, it can treat flexibly linear and nonlinear data, this paper proposes an adaptive DDoS attack detection method based on the ensemble learning framework.

**4.1 The multiple-kernel learning model**

The multiple-kernel learning (MKL) model is developed from the original single- kernel SVM. In single- kernel SVM, a SVM only uses one kernel function to map the sample to high dimensional spaces. By comparison, the multiple-kernel learning model uses multiple-kernel functions with weight to map the sample to high-dimensional space. Therefore, it has higher flexibility and adaptability on heterogeneous data.

The multiple kernel learning is defined as follows: given training set $T = \{(x_1, y_1), (x_2, y_2) \cdots (x_n, y_n)\}$, testing set $C = \{x_1^{'}, x_2^{'} \cdots x_s^{'}\}$, $x_i \in R^d$, $x_k^{'} \in R^d$, $y_i \in (-1, +1)$, R is real-number set, d is data dimension, $i = 1, 2, \cdots, n$, $k = 1, 2, \cdots s$. $K_1(x, x^{'}), K_2(x, x^{'}) \cdots K_M(x, x^{'})$ are kernel functions in $R^d \times R^d$, $\phi_1, \phi_2 \cdots \phi_M$ is a kernel mapping for each function. In the classic multiple-kernel learning SimpleMKL [53], the objective function of the hyperplane is as follows:

$$f(x) = \sum_{m=1}^{M}(\omega_m, \phi_m(x)) + b \quad (11)$$

where $\omega_m$ is the weight for each kernel function, and b is bias. The relaxation factor is $\xi$. According to the principle of minimum structure, the objective function can be optimized as follows:

$$\min \psi(\omega_m, b, \xi, d) = \frac{1}{2}\sum_{m=1}^{M}\frac{1}{d_m}\|\omega_m\|_{H_m}^2 + C\sum_{i=1}^{n}\xi_i \quad (12)$$

s.t.

$$\begin{cases} y_i \sum_{m=1}^{M}\omega_m \cdot \varphi(x_i) + y_i b \geq 1 - \xi_i \\ \sum_{m=1}^{M}d_m = 1, d_m \geq 0 \\ \xi_i \geq 0 \end{cases} \quad (13)$$

By the two-order alternation optimization, the formula (12) can be converted to the optimization problem with $d_m$ as the variable:

$$\min_{d \geq 0} J(d), \sum_{m=1}^{M}d_m = 1 \quad (14)$$

s.t.

$$\begin{cases} \min_{\omega_m, b, \xi} = \frac{1}{2}\sum_{m=1}^{M}\frac{1}{d_m}\|\omega_m\|_{H_m}^2 + C\sum_{i=1}^{n}\xi_i \\ y_i \sum_{m=1}^{M}\omega_m \cdot \varphi(x_i) + y_i b \geq 1 - \xi_i \\ \xi_i \geq 0 \end{cases} \quad (15).$$

The Lagrange function of $J(d)$ is as follows:

$$L = \frac{1}{2}\sum_{m=1}^{M}\frac{1}{d_m}\|\omega_m\|_{H_m}^2 + C\sum_{i=1}^{n}\xi_i + \sum_{i=1}^{m}\alpha_i(1-\xi_i-y_i\sum_{m=1}^{M}\omega_m \cdot \varphi_m(x_i) + y_i b) + \sum_{i=1}^{n}v_i\xi_i \quad (16)$$

where $\alpha_i, v_i$ are Lagrange operators. First, $\omega_m, b, \xi_i$ are calculated for partial derivatives. Then, the extremums are gained when the partial derivatives are "0." Finally, extremums are brought into the Lagrange function, which can be further changed to:

$$\max Q(\alpha) = -\frac{1}{2}\sum_{i,j=1}^{m}\alpha_i\alpha_j y_i y_j K_d(x_i, x_j) + \sum_{i=1}^{n}\alpha_i \quad (17)$$

s.t.

$$\begin{cases} \sum_{i}^{n} \alpha_i y_i = 0 \\ C \geq \alpha_i \geq 0 \\ K_d(x_i, x_j) = \sum_{m=1}^{M} d_m k_m(x_i, x_j) \end{cases} \quad (18)$$

The gradient descent method is used to adjust $J(d)$ on $d$, update $d$, and optimize the $d$ as well as $a$ alternately. Then, an optimal solution is obtained:

$\alpha^* = (\alpha_1, \alpha_2, \cdots, \alpha_n)$; that is, the original objective function eventually turns into (19). The detailed formulation is as follows:

$$f(x) = \sum_{i=1}^{n} \alpha_i^* y_i \sum_{m=1}^{M} d_m K_d(x_i, x_j) + b \quad (19)$$

$x_j \in C$. When the test set data as $x_j$ is inputted to $f(x)$, the object function can determine the category of test set data.

### 4.2 The attack detection model based on multiple-kernel learning

The SimpleMKL model can be suitable for the all dimension weight values with "1". But it cannot fully exert the different features. This paper uses the feature weights to control the effect of different features on the model. To gain the appropriate feature weights in the SimpleMKL model, we combine the gradient method to optimize the weight parameters, so that the detection accuracy is further improved.

We marked ACD as $x_1$, IBF as $x_2$, MFF as $x_3$, HIAD as $x_4$, and FFV as $x_5$, then the feature value vector is $F = (x_1, x_2, x_3, x_4, x_5)$, and the marked weight vector is $W = (w_1, w_2, w_3, w_4, w_5)$. Combinatorial features are $CF = F * W^T$, and the mean value of each dimension of normal flow is $u_{11}$, $u_{12}$, $u_{13}$, $u_{14}$, or $u_{15}$. Note the mean value of each dimension of the attack flow is $u_{21}$, $u_{22}$, $u_{23}$, $u_{24}$, or $u_{25}$.

The inter-class mean squared difference is expressed as follows:

$$M = [w_1 * (u_{11} - u_{21})]^2 + [w_2 * (u_{12} - u_{22})]^2 + [w_3 * (u_{13} - u_{23})]^2 + [w_4 * (u_{14} - u_{24})]^2 + [w_5 * (u_{15} - u_{25})]^2$$

The normal intra-class variance is denoted:

$$S_1 = \sum_{i=1}^{n} [w_1 * (x_{i1} - u_{11})]^2 + [w_2 * (x_{i2} - u_{12})]^2 + [w_3 * (x_{i3} - u_{13})]^2 + [w_4 * (x_{i4} - u_{14})]^2 + [w_5 * (x_{i5} - u_{15})]^2$$

The attack intra-class variance is denoted:

$$S_2 = \sum_{i=1}^{n} [w_1 * (x_{i1} - u_{21})]^2 + [w_2 * (x_{i2} - u_{22})]^2 + [w_3 * (x_{i3} - u_{23})]^2 + [w_4 * (x_{i4} - u_{24})]^2 + [w_5 * (x_{i5} - u_{25})]^2$$

The intra-class variance is $S = S_1 + S_2$. To improve classification accuracy and ensure a rapid convergence of functions, on the one hand, we should try to improve the mean difference between positive and negative samples, so that the two kinds of samples are far away from each other, that is, we should increase the M value. On the other hand, we should minimize the differences between samples. The variance corresponding to each dimension should be as small as possible, thus reducing the S value. Therefore, the classification model needs to train two different classifiers to classify the samples. One classifier is inter-class mean squared difference growth (M-SMKL) and the other classifier is intra-class variance descent (S-SMKL). In combination with the SimpleMKL framework formula (12), the above problems can be transformed into (20). The detailed formulation is as follows:

$$\begin{cases} \max_{x_{ij} \in F} \alpha M + \min_{x_{ij} \in F} \beta S \\ \min \psi(\omega_m, b, \xi, d, w) = \frac{1}{2}\sum_{m=1}^{M} \frac{1}{d_m} \|\omega_m\|_{H_m}^2 + C\sum_{i=1}^{n} \xi_i \end{cases} \quad (20)$$

s.t.

$$\begin{cases} y_i \sum_{m=1}^{M} \omega_m \cdot \varphi(wx_i) + y_i b \geq 1 - \xi_i \\ \sum_{m=1}^{M} d_m = 1, d_m \geq 0 \\ \xi_i \geq 0 \\ M < \sigma_1 \\ S > \sigma_2 \end{cases} \quad (21)$$

If $\alpha \gg \beta$, the objective function is M-SMKL. If $\beta \gg \alpha$, the objective function is S-SMKL. $\alpha$ and $\beta$ are converted to the learning rate of formula (29).

To solve the above problems, we use the way of iterative updating weights to get the objective function. The details are as follows. Firstly, the weights of each feature are assigned initial values. Secondly, combine with the (20) and (21) to gain optimal function of this time. The mathematical form is expressed as follows:

$$\max Q(\alpha) = -\frac{1}{2}\sum_{i,j=1}^{m} \alpha_i \alpha_j y_i y_j K_d(wx_i, wx_j) + \sum_{i=1}^{n} \alpha_i \quad (22)$$

s.t.

$$\begin{cases} \sum_{i}^{n} \alpha_i y_i = 0 \\ C \geq \alpha_i \geq 0 \\ K_d(wx_i, wx_j) = \sum_{m=1}^{M} d_m k_m(wx_i, wx_j) \end{cases}$$

(23)

The optimal equation obtained using the above equations (22) and (23) is as follows:

$$f(x) = \sum_{i=1}^{n} \alpha_i^* y_i \sum_{m=1}^{M} d_m K_d(wx_i, wx_j) + b . \quad (24)$$

To further determine whether the optimal equation has achieved good results, this paper sets two constraint conditions for M-SMKL and S-SMKL respectively without conflict with the formula (21) constraint conditions. These constraint conditions is expressed as follows:

The constraint conditions of M-SMKL are as follows:

$$\begin{cases} t_1 < |M_{i+1} - M_i| < t_2 < |M_i - M_{i-1}| < t_3 \\ \dfrac{f(M_i)}{f(S_i)} - \dfrac{f(M_{i-1})}{f(S_{i-1})} > p_1 \\ \dfrac{f(M_i)}{f(S_i)} - \dfrac{f(M_{i+1})}{f(S_{i+1})} > p_2 \end{cases} \quad (25).$$

The constraint conditions of S-SMKL are as follows:

$$\begin{cases} t_4 < |S_i - S_{i-1}| < t_5 < |S_{i+1} - S_i| < t_6 \\ \dfrac{f(M_i)}{f(S_i)} - \dfrac{f(M_{i-1})}{f(S_{i-1})} > p_3 \\ \dfrac{f(M_i)}{f(S_i)} - \dfrac{f(M_{i+1})}{f(S_{i+1})} > p_4 \end{cases} \quad (26).$$

where the values of $p_1$, $p_2$, $p_3$, $p_4$ are close to "0"; the values of $t_1$, $t_2$ and $t_3$ are close to "1"; the values of $t_4$, $t_5$, $t_6$ are close to "7.5". If the constraint condition is satisfied, the algorithm will be stopped and the formula (24) will become the optimal function, otherwise, the each dimension weight will be updated iteratively. The gradient of M and S corresponding to the each dimension weight is as follows:

$$\begin{cases} \dfrac{\partial M}{\partial w_1}=2w_1(u_{11}-u_{21})^2 \\ \dfrac{\partial M}{\partial w_2}=2w_2(u_{12}-u_{22})^2 \\ \dfrac{\partial M}{\partial w_3}=2w_3(u_{13}-u_{23})^2 \\ \dfrac{\partial M}{\partial w_4}=2w_4(u_{14}-u_{24})^2 \\ \dfrac{\partial M}{\partial w_5}=2w_5(u_{15}-u_{25})^2 \end{cases} \quad (27)$$

$$\begin{cases} \dfrac{\partial S}{\partial w_1}=2\left[w_1(\sum_{i=1}^{n_1}x_{11}^2-n_1u_{11}^2)+w_1(\sum_{i=1}^{n_2}x_{21}^2-n_2u_{21}^2)\right] \\ \dfrac{\partial S}{\partial w_2}=2\left[w_2(\sum_{i=1}^{n_1}x_{12}^2-n_1u_{12}^2)+w_2(\sum_{i=1}^{n_2}x_{22}^2-n_2u_{22}^2)\right] \\ \dfrac{\partial S}{\partial w_3}=2\left[w_3(\sum_{i=1}^{n_1}x_{13}^2-n_1u_{13}^2)+w_3(\sum_{i=1}^{n_2}x_{23}^2-n_2u_{23}^2)\right] \\ \dfrac{\partial S}{\partial w_4}=2\left[w_4(\sum_{i=1}^{n_1}x_{14}^2-n_1u_{14}^2)+w_4(\sum_{i=1}^{n_2}x_{24}^2-n_2u_{24}^2)\right] \\ \dfrac{\partial S}{\partial w_5}=2\left[w_5(\sum_{i=1}^{n_1}x_{15}^2-n_1u_{15}^2)+w_5(\sum_{i=1}^{n_2}x_{25}^2-n_2u_{25}^2)\right] \end{cases} \quad (28)$$

where $n_1$ is the number of the normal flow feature of the training sample; $n_2$ is the number of the attack flow feature of the training sample. According to gradients in equations (27) and (28), the weight of each dimension is updated as follows (29):

$$\begin{cases} w_1 = w_1 + 2*lr_1*\dfrac{\partial M}{\partial w_1} - 2*lr_2*\dfrac{\partial S}{\partial w_1} \\ w_2 = w_2 + 2*lr_1*\dfrac{\partial M}{\partial w_2} - 2*lr_2*\dfrac{\partial S}{\partial w_2} \\ w_3 = w_3 + 2*lr_1*\dfrac{\partial M}{\partial w_3} - 2*lr_2*\dfrac{\partial S}{\partial w_3} \\ w_4 = w_4 + 2*lr_4*\dfrac{\partial M}{\partial w_4} - 2*lr_2*\dfrac{\partial S}{\partial w_4} \\ w_5 = w_5 + 2*lr_1*\dfrac{\partial M}{\partial w_5} - 2*lr_2*\dfrac{\partial S}{\partial w_5} \end{cases} \quad (29)$$

where $lr_1$ is the learning rate of gradient ascent; $lr_2$ is the learning rate of gradient descent. $lr_1$ has the same function as $\alpha$ and $lr_2$ has the same function as $\beta$. Each weight of the updated is multiplied by each original feature accordingly and the next round of iteration is carried out

### 4.3 Framework of multiple-kernel learning detection based on ensemble learning

We input the multidimensional data with weight and set the learning rate. Then two different classifiers are trained. M-SMKL is trained by increasing the M value mainly with reducing the S value secondarily and the S-SMKL is trained by reducing the S value mainly with increasing the M value secondarily. During the training process, the M value and the S value are constantly updated with the method of gradient rising and descending until the constraint conditions are met. The flowchart is provided in Figure 1.

The detection process is as follows: firstly, the test data is multiplied with two different weight vector which are trained earlier; secondly, the calculated data are inputted to the corresponding M-SMKL and S-SMKL model; finally, use the sliding window mechanism to coordinate two kinds of models. The sliding window mechanism is described as follows. Firstly, a sliding window with a size of *n* is created. Secondly, the trained M-SMKL classifies the test data and obtain the first classification results; the trained S-SMKL classifies the test data and obtain the second classification results. Finally, four kinds of ways are used to cooperatively treat the first classification results and the second classification results, the details are as follows: (1) if M-SMKL and S-SMKL identify that the current data category is both normal, the current data category is judged to be normal; (2) if M-SMKL and S-SMKL identify that the current data category is both attack, the current data category is judged to be attack; (3) if M-SMKL identify that the current data category is normal but S-SMKL identify that the current data category is attack, the current data category is judged to be attack; (4) if M-SMKL identify that the current data category is attack but S-SMKL identify that the current data category is normal, then, step 1. move the starting point of the sliding window to the current position of the test data in the first classification result, and map the end point of the sliding window to the n-1 position of the first classification results; step 2. if the results in the sliding window are all attack, the current data category is judged to be attack, otherwise, the current data category is judged to be normal. The flow chart is provided in Figure 2.

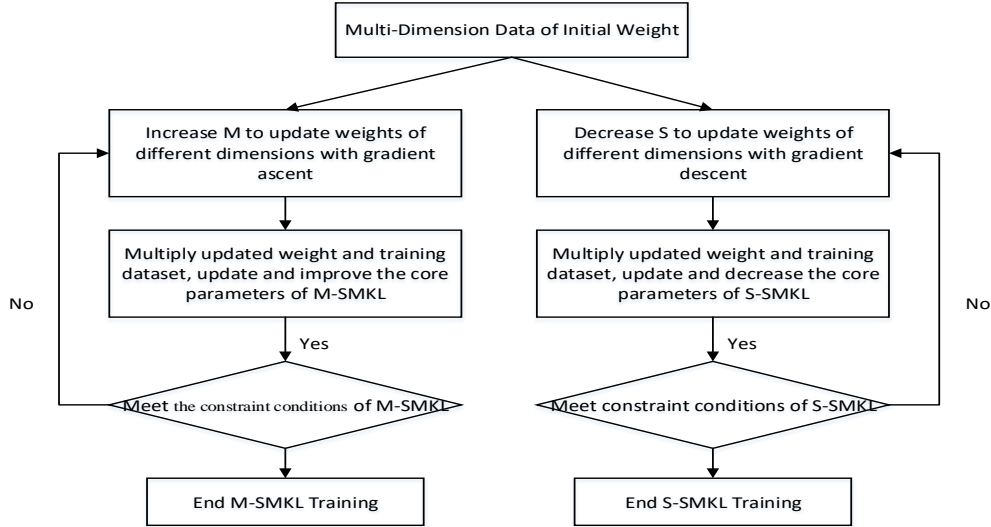

Figure 1: Flow chart of multiple-kernel learning training process based on ensemble Learning

The reason for the training of two kinds of SMKL is that S-SMKL focuses on reducing the difference between the data of each dimension and can assemble the two types of samples in their respective central positions. However, S-SMKL does not consider the location of the two sample-center points. Although a better classification feature can be maintained on the whole, it is impossible to identify DDoS attacks earlier because of center distance of the normal flow and attack flow is small. M-SMKL focuses on the difference between the two types of data centers and maximizes the sample centers distance between the two types of sample centers, making the two samples as separate as possible. M-SMKL can expand the distance of different class so that the attack flow can be identified earlier but it makes intra-class data dispersed, causing default results. Therefore, the sliding window mechanism is adopted to coordinate the two models to detect early DDoS accurately.

## 5. Experimental analysis

### 5.1 Experimental Data Sets and Evaluation Standards

The data set used for this experiment is the CAIDA "DDoS Attack 2007" data set [54]. This data set contains an [L1] Distributed Denial of Service (DDoS) anonymous traffic attack for approximately one hour on August 4, 2007. The total size of the data set is 21 GB, which accounts for approximately one hour (20:50:08 UTC–21:56:16 UTC). Attacks began around 21:13, causing the network load to grow rapidly (in minutes) from approximately 200 kbits/s to 80 megabits/s. One hour of attack traffic is divided into 5 minutes of files and stored in PCAP format. The contents of this data set are TCP network traffic packets. Each TCP packet contains the source address, destination address, source port, destination port, packet size, and protocol type. The duration of normal flow data used in this paper is 2 minutes in total, and the duration of attack data is 5 minutes in total.

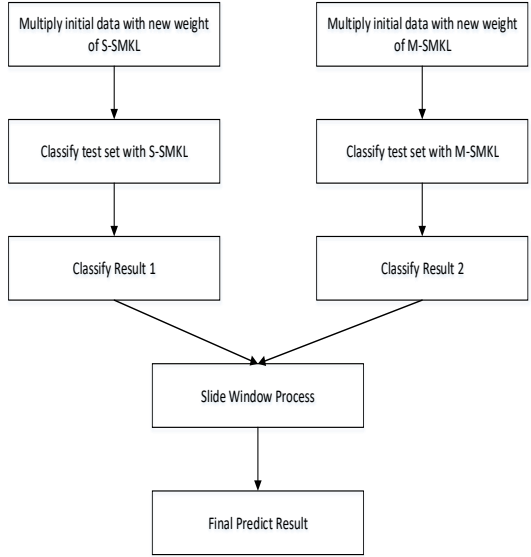

Figure 2: Flow chart of multiple kernel learning detection process based on ensemble learning.

The hardware equipment adopted is 8 GB memory, Intel Core i7 processor and a computer with a Windows 10 64-bit system; the development environment is MATLAB 2014a and Codeblocks 10.05. The evaluation criteria used in this paper consist of the detection rate (DR), the false alarm rate (FR), and total error rate (ER).

Assume that TP indicates that the number of normal test samples is properly marked, FP indicates the number of normal test samples that have been incorrectly marked, TN indicates the number of attack test samples that are correctly marked, and FN indicates the number of attack test samples that have been incorrectly marked:

$$\begin{cases} DR = \dfrac{TN}{TN + FN} \\ FR = \dfrac{FP}{TP + FP} \\ ER = \dfrac{FN + FP}{TP + FP + TN + FN} \end{cases}. \quad (30)$$

We used the above five feature extraction algorithms to extract features from the data set. The extracted feature values are normalized and used as a training set. The data in the training set can be regarded as the regularity of the change in network traffic. The network traffic has an abrupt and volatile nature. Therefore, although the collected network data have similarities with the conventional ones, they still have a certain degree of difference. To simulate this phenomenon for verifying the effectiveness of the presented method, three types of data are generated as follows: (1) Each normal flow feature values and attack flow feature values are multiplied by random number; (2) only the attack flow feature values are multiplied by random number; (3) only the normal flow feature values are multiplied by random number.

### 5.2 Experimental Results and Analysis

Five features are used to extract feature data from attack data and normal data, and positive as well as negative sample sets are obtained. The sampling time is set to 1 s, and the remaining parameters of the five feature extraction methods are set as follow: $\theta_1 = 0.5$, $\theta_2 = 0.5$, $\theta_3 = 3$, $\theta_4 = 3$, $\theta_5 = 3$, $\theta_6 = 3$, $\theta_7 = 3$, $\theta_8 = 3$, and $\theta_9 = 3$. A total of normal feature values is 211 and a total of attack feature values is 280. Figure 3-9 illustrates the feature values extracted by the five algorithms.

As illustrated in Figure 3, the early attack feature values of DDoS attack are close to the normal feature values. This is because there are a large number of bidirectional flows in the early stage of the DDoS attack and these bidirectional flows gradually decrease with the increase of the attack degree. Therefore, using the ACD as a feature after 70 seconds can significantly reflect the difference between the attack flow and the normal flow. ACD can be able to reflect the difference between normal flow and attack flow earliest.

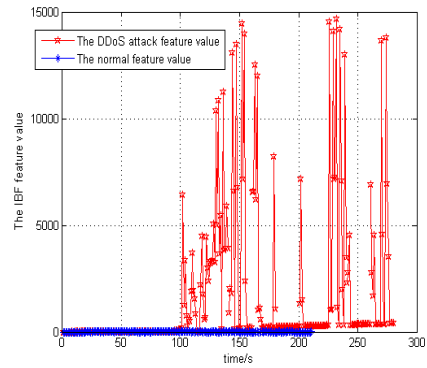

Figure 4: The IBF feature graph of DDoS attack flow and normal flow.

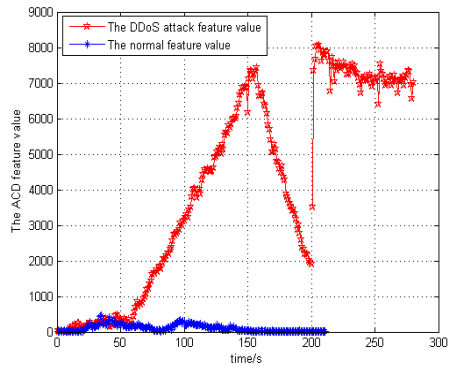

Figure 3: The ACD feature graph of DDoS attack flow and normal flow.

As illustrated in Figure 4, compared with ACD, although IBF does not recognize the attack flow earlier, the distribution range of its feature values is more uniform and presents a certain degree of volatility. This makes the feature less susceptible to individual outliers.

As illustrated in Figure 5, the FFV feature is very similar to the ACD, but as illustrated in Figure. 6 and Figure 7, in the initial stage, the FFV is more capable of reflecting the difference between the attack flow and the normal flow than the ACD is.

As illustrated in Figure 8, although the MFF feature cannot determine the attack flow and the normal flow as early as possible, it can make the feature values of the attack stage more stable, so that it can avoid the outliers of attack flows. .

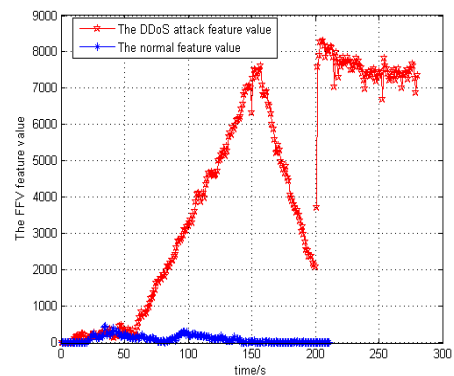

Figure 5: The FFV feature graph of DDoS attack flow and normal flow.

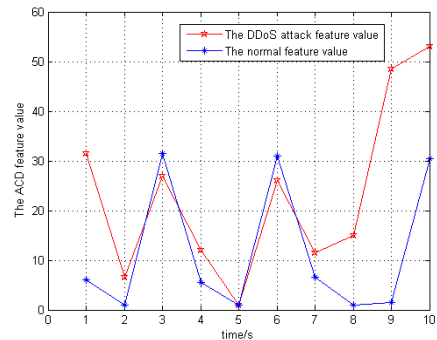

Figure 6: The ACD feature graph of DDoS attack flow and normal flow in the first 10 seconds.

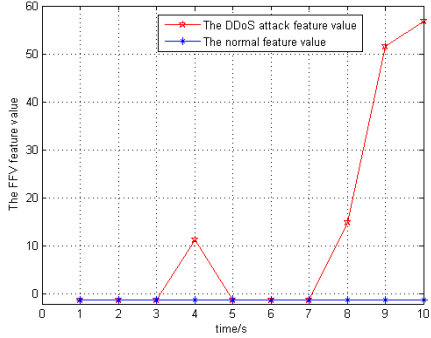

Figure 7: The FFV feature graph of DDoS attack flow and normal flow in the first 10 seconds.

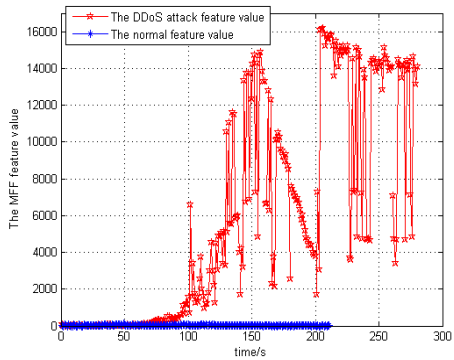

Figure 8: The MFF feature graph of DDoS attack flow and normal flow.

As illustrated in Figure 9, it can be seen from the value of the ordinate that the HIAD best reflects the difference between the normal flow and the attack flow while having better stability in the latter half of the attack flow. After the early data, this feature can greatly distinguish between normal flow and abnormal flow, influence the classifier more and make better decisions.

In summary, all five features have their own unique characteristics. To make full use of the characteristics of each feature, the feature values extracted by these five algorithms are each used as a five-dimensional-feature data set. Using these five feature values as training sets, two multiple kernel learning models dominated by gradient ascent and gradient descent are trained into the algorithm, and corresponding five-dimensional feature weight vectors are obtained. Finally, according to the framework of figure 2, the classification results of test set are obtained and are used to verify the effectiveness of method. The parameters of M-SMKL are set as follow: $l_{r_1} = 2*10^{-5}$, $l_{r_2} = 2*10^{-3}$, $t_1 = 1.002$, $t_2 = 1.0065$, $t_3 = 1.007$, $p_1 = 0.000084$, and $p_2 = 0.000001$. The parameters of S-SMKL are set as follow: $l_{r_1} = 2*10^{-5}$, $l_{r_2} = 2*10^{-2}$, $t_4 = 7.3425$, $t_5 = 7.8340$, $t_6 = 7.8350$, $p_3 = 0.000775$, and $p_4 = 0.000680$. The size of the sliding window is 8. The parameters for multiple-kernel learning are all default values, and the kernel function includes two Gaussian functions and two polynomial functions. The SVM parameters are all default values, and the kernel function is linear function. The experimental results are illustrated in Figures 10–18.

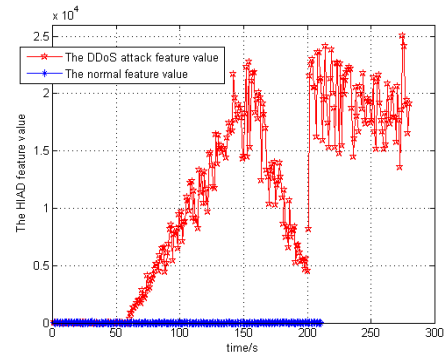

Figure 9: The HIAD feature graph of DDoS attack flow and normal flow.

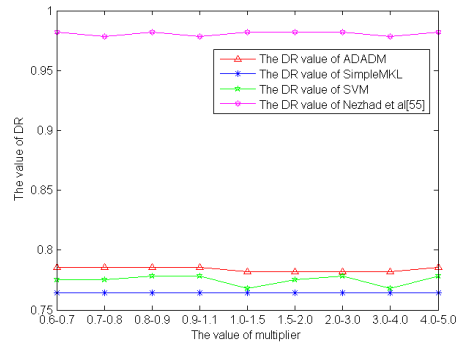

Figure 10: The DR contrast diagram of four algorithms for scaling attack flow and normal flow.

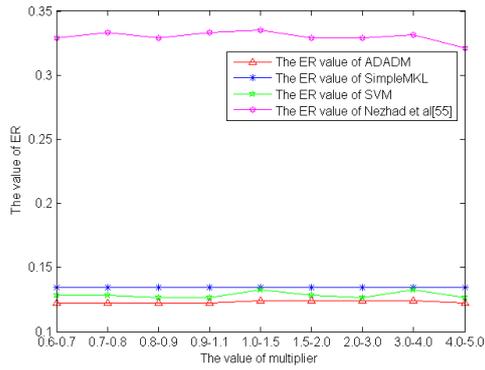

Figure 11: The ER contrast diagram of four algorithms for scaling attack flow and normal flow.

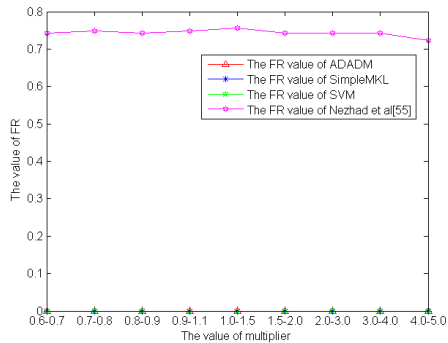

Figure 12: The FR contrast diagram of four algorithms for scaling attack flow and normal flow.

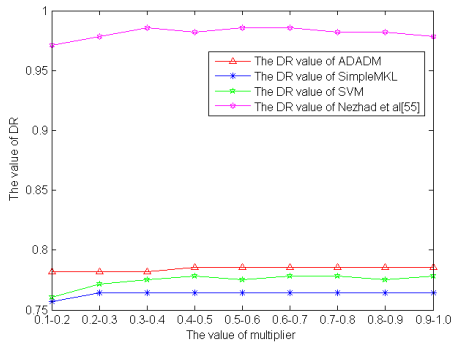

Figure 13: The DR contrast diagram of four algorithms for narrowing the attack flow.

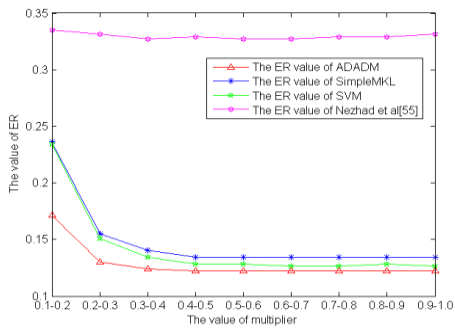

Figure 14: The ER contrast diagram of four algorithms for narrowing the attack flow.

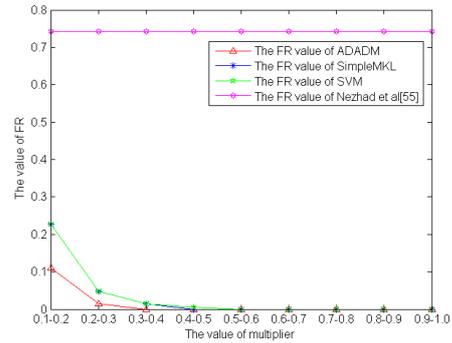

Figure 15: The FR contrast diagram of four algorithms for narrowing the attack flow.

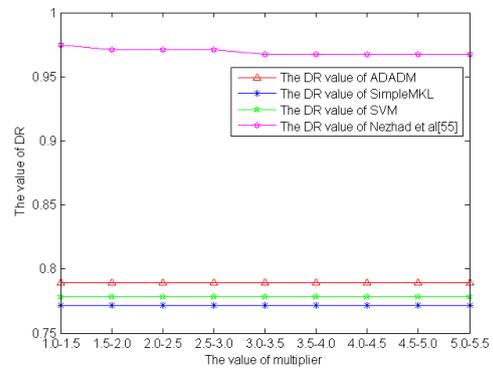

Figure 16: The DR contrast diagram of four algorithms for amplifying the normal flow.

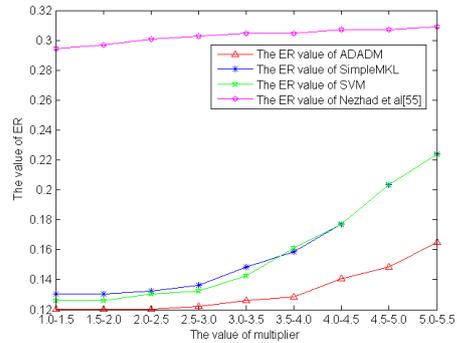

Figure 17: The ER contrast diagram of four algorithms for amplifying the normal flow.

As shown in figure 10-18, under the three types of experiments, according to the three evaluation criteria, the overall performance of the algorithms from the highest to the lowest is the method of the ADADM, the SVM method, the SMKL method and Nezhad et al's method [55].

This is because although the method described by Nezhad et al. [55] is visibly superior to other methods in terms of DR indicators, it is far worse than other methods with respect to other indicators. The reason is that the Nezhad et al [55] method relies excessively on the first reference point. When the first reference point fluctuates, this method recognizes easily some normal samples as attack samples.

Although the classification accuracy of the attack samples is high, a large number of normal samples are misjudged, so this method is superior in terms of DR and its other indicators are inferior to those of other methods. This is why, in this case, the Nezhad et al [55]'s method performs the worst. The effect of SVM is generally better than that of the SMKL method because although the SMKL method coordinates multiple kernel functions to map the sample to a high-dimensional Hilbert space, the linear kernel function is obviously more suitable for the sample. Using the linear kernel SVM can establish a better hyperplane than the SMKL method to identify the data containing early DDoS attacks. However, although the multiple-kernel learning method does not use a linear kernel function that is more suitable for the sample space, it can still maintain high accuracy, indicating that multiple-kernel learning has a lower dependence on the selection of kernel functions than the single kernel SVM.

We compared the ADADM to the SVM method. The ADADM method uses the same kernel function as SMKL method. Because the multi-kernel learning method is flexible and adaptable, it is possible to continuously optimize the hyperplane by adjusting the weights of the feature of each dimension to recognize the DDoS as early as possible. Attack flow data and normal flow data are located on both sides of the hyperplane.

Table 1: Comparison results of four algorithms for scaling attack flow and normal flow

|  |  | The value of the random multiplier | | | | | | | | |
| --- | --- | --- | --- | --- | --- | --- | --- | --- | --- | --- |
|  |  | 0.6–0.7 | 0.7–0.8 | 0.8–0.9 | 0.9–1.1 | 1.0–1.5 | 1.5–2.0 | 2.0–3.0 | 3.0–4.0 | 4.0–5.0 |
| ADADM method | DR (%) | 78.57 | 78.57 | 78.57 | 78.57 | 78.21 | 78.21 | 78.21 | 78.21 | 78.57 |
|  | FR (%) | 0.01 | 0.01 | 0.01 | 0.01 | 0.01 | 0.01 | 0.01 | 0.01 | 0.01 |
|  | ER (%) | 12.22 | 12.22 | 12.22 | 12.22 | 12.42 | 12.42 | 12.42 | 12.42 | 12.22 |
| SimpleMKL method | DR (%) | 76.43 | 76.43 | 76.43 | 76.43 | 76.43 | 76.43 | 76.43 | 76.43 | 76.43 |
|  | FR (%) | 0.01 | 0.01 | 0.01 | 0.01 | 0.01 | 0.01 | 0.01 | 0.01 | 0.01 |
|  | ER (%) | 13.44 | 13.44 | 13.44 | 13.44 | 13.44 | 13.44 | 13.44 | 13.44 | 13.44 |
| SVM method | DR (%) | 77.50 | 77.50 | 77.86 | 77.86 | 76.79 | 77.50 | 77.86 | 76.79 | 77.86 |
|  | FR (%) | 0.01 | 0.01 | 0.01 | 0.01 | 0.01 | 0.01 | 0.01 | 0.01 | 0.01 |
|  | ER (%) | 12.83 | 12.83 | 12.63 | 12.63 | 13.24 | 12.83 | 12.63 | 13.24 | 12.63 |
| Nezhad et al [55]'s method | DR (%) | 98.21 | 97.85 | 98.21 | 97.85 | 98.21 | 98.21 | 98.21 | 97.85 | 98.21 |
|  | FR (%) | 74.29 | 74.76 | 74.29 | 74.76 | 75.71 | 74.29 | 74.29 | 74.29 | 72.38 |
|  | ER (%) | 32.92 | 33.33 | 32.92 | 33.33 | 33.54 | 32.92 | 32.92 | 33.13 | 32.11 |

In addition, using the idea of ensemble learning to train two different classifiers and using the sliding window mechanism to further synthesize the advantage of each classifier improves the algorithm's performance in the three types of experiments. This method we propose outperforms not only the SVM method but also other methods of DDoS attack detection. The experimental data are presented in Table 1, Table 2, and Table 3.

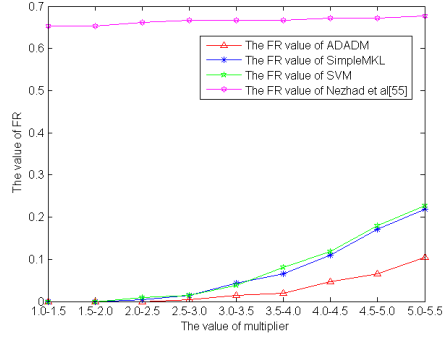

Figure 18: The FR contrast diagram of four algorithms for amplifying the normal flow

Table 2: Comparison results of four algorithms for narrowing the attack flow

| | | The value of the random multiplier | | | | | | | | |
|---|---|---|---|---|---|---|---|---|---|---|
| | | 0.1–0.2 | 0.2–0.3 | 0.3–0.4 | 0.4–0.5 | 0.5–0.6 | 0.6–0.7 | 0.7–0.8 | 0.8–0.9 | 0.9–1.0 |
| ADADM method | DR (%) | 78.21 | 78.21 | 78.21 | 78.57 | 78.57 | 78.57 | 78.57 | 78.57 | 78.57 |
| | FR (%) | 10.99 | 1.42 | 0.01 | 0.01 | 0.01 | 0.01 | 0.01 | 0.01 | 0.01 |
| | ER (%) | 17.15 | 13.04 | 12.42 | 12.22 | 12.22 | 12.22 | 12.22 | 12.22 | 12.22 |
| SimpleMKL method | DR (%) | 75.71 | 76.43 | 76.43 | 76.43 | 76.43 | 76.43 | 76.43 | 76.43 | 76.43 |
| | FR (%) | 22.75 | 4.74 | 1.42 | 0.01 | 0.01 | 0.01 | 0.01 | 0.01 | 0.01 |
| | ER (%) | 23.63 | 15.48 | 14.05 | 13.44 | 13.44 | 13.44 | 13.44 | 13.44 | 13.44 |
| SVM method | DR (%) | 76.07 | 77.14 | 77.50 | 77.86 | 77.50 | 77.86 | 77.86 | 77.50 | 77.86 |
| | FR(%) | 22.75 | 4.74 | 1.42 | 0.47 | 0.01 | 0.01 | 0.01 | 0.01 | 0.01 |
| | ER(%) | 23.42 | 15.07 | 13.44 | 12.83 | 12.83 | 12.63 | 12.63 | 12.83 | 12.63 |
| Nezhad et al [55]'s method | DR(%) | 97.13 | 97.85 | 98.57 | 98.21 | 98.57 | 98.57 | 98.21 | 98.21 | 97.85 |
| | FR(%) | 74.29 | 74.29 | 74.29 | 74.29 | 74.29 | 74.29 | 74.29 | 74.29 | 74.29 |
| | ER (%) | 33.54 | 33.13 | 32.72 | 32.92 | 32.72 | 32.72 | 32.92 | 32.92 | 33.13 |

## 6. Conclusion

In this paper, five-dimensional features are defined for describing the burstiness of DDoS attack flows, the distribution of IP source addresses and the interactivity of DDoS attack flows. Based on the five-dimensional features and the ensemble learning framework, adaptive feature weights are obtained and the M-SMKL and S-SMKL multiple kernel learning models are trained to detect DDoS attack. For identifying early attacks effectively, the sliding window mechanism is used to coordinate the S-SMKL and the M-SMKL to deal with the detection results.

Experimental results show that, compared with similar methods, our method can produce more accurate results for detecting early DDoS attack. We believe that the approach will have great value in the security of cloud computing, cloud robotics [56], intelligent transportation [57], IOT and so on.

In the follow-up work, we will further study how to transform the multi-dimensional weight adaptive problem based on multiple kernel learning into a convex optimization problem, and improve the detection rate and convergence speed of the method.

Table 3: Comparison results of four algorithms for amplifying the normal flow

|  |  | The value of random multiplier | | | | | | | | |
|---|---|---|---|---|---|---|---|---|---|---|
|  |  | 1.0–1.5 | 1.5–2.0 | 2.0–2.5 | 2.5–3.0 | 3.0–3.5 | 3.5–4.0 | 4.0–4.5 | 4.5–5.0 | 5.0–5.5 |
| ADADM method | DR (%) | 78.93 | 78.93 | 78.93 | 78.93 | 78.93 | 78.93 | 78.93 | 78.93 | 78.93 |
|  | FR (%) | 0.01 | 0.01 | 0.01 | 0.47 | 1.42 | 1.90 | 4.74 | 6.64 | 10.43 |
|  | ER (%) | 12.02 | 12.02 | 12.02 | 12.22 | 12.63 | 12.83 | 14.05 | 14.87 | 16.50 |
| SimpleMKL method | DR (%) | 77.14 | 77.14 | 77.14 | 77.14 | 77.14 | 77.14 | 77.14 | 77.14 | 77.14 |
|  | FR (%) | 0.01 | 0.01 | 0.47 | 1.42 | 4.27 | 6.64 | 10.90 | 17.06 | 21.80 |
|  | ER (%) | 13.04 | 13.04 | 13.24 | 13.65 | 14.87 | 15.89 | 17.72 | 20.37 | 22.40 |
| SVM method | DR (%) | 77.86 | 77.86 | 77.86 | 77.86 | 77.86 | 77.86 | 77.86 | 77.86 | 77.86 |
|  | FR (%) | 0.01 | 0.01 | 0.95 | 1.42 | 3.79 | 8.06 | 11.85 | 18.01 | 22.75 |
|  | ER (%) | 12.63 | 12.63 | 13.04 | 13.24 | 14.26 | 16.09 | 17.72 | 20.37 | 22.40 |
| Nezhad et al [55]'s method | DR (%) | 97.49 | 97.13 | 97.13 | 97.13 | 96.77 | 96.77 | 96.77 | 96.77 | 96.77 |
|  | FR (%) | 65.24 | 65.24 | 66.19 | 66.67 | 66.67 | 66.67 | 67.14 | 67.14 | 67.62 |
|  | ER (%) | 29.47 | 29.67 | 30.08 | 30.28 | 30.49 | 30.49 | 30.69 | 30.69 | 30.90 |

## Conflicts of Interest

There are no conflicts of interest in this paper.

## Acknowledgment:

This work was supported by the Hainan Provincial Natural Science Foundation of China [2018CXTD333, 617048]; The National Natural Science Foundation of China [ 61762033, 61702539]; Hainan University Doctor Start Fund Project [kyqd1328]; Hainan University Youth Fund Project [qnjj1444]..